\documentclass[twoside]{aiml22}

\usepackage{aiml22macro}

\usepackage{graphicx}
\usepackage{amsmath}
\usepackage{amssymb}
\usepackage{stmaryrd}
\usepackage{mathptmx}
\usepackage{wasysym}
\usepackage{inputenc}

\def\lastname{Khaitovich}

\begin{document}

\begin{frontmatter}
  \title{Neighbourhood semantics  and axioms for strategic fragment of classical stit logic}
  \author{Daniil Khaitovich}\footnote{dkhaytovich@hse.ru}
  \address{International Laboratory for Logic, Linguistics and Formal Philosophy, HSE \\ Moscow, Russia}

  \begin{abstract}
  STIT (sees to it that) semantics is one of the most prominent tools in modal logic of agency, widely used among both philosophers and responsible AI scholars. STIT logic surveys the properties of agents seeing to it that some state of affairs holds without specifying concrete actions by which that state of affairs is guaranteed. In comparison with other multi-agent modal logics, the main advantage of STIT theories is expressive power. STIT logic allows to study not only statements about agents' abilities to perform certain actions (as it is in variations of Coalition Logic or Propositional Dynamic Logic), but about what  choices they make and what they de-facto achieve as well. 

  Nevertheless, in some occasions such expressivity may be redundant. This paper surveys a specific fragment of classical STIT logic, which has only strategic modal operator $[i]\phi$, what stands for the fact that agent $i$ has an ability to see to it that $\phi$ holds. The neighbourhood semantics for the fragment is presented, accompanied with the soundness, canoniciy hence strong completeness results. Furthermore, the paper presents basic considerations on epistemic extension of the presented fragment. 
  \end{abstract}

 \end{frontmatter}

\section{Introduction}
The stit (abbreviation for \textit{\textquotedblleft sees to it that\textquotedblright}) theory is one of the most prominent modal theories of agency. In stit framework, actions of individuals are studied via sentences of the form \textquotedblleft agent $i$ sees to it that $\phi$ holds\textquotedblright, where the syntactic construction \textquotedblleft agent $i$ sees to it that\textquotedblright $\:$  is treated as a modality $[i: stit]$ (we will use a simpler notation: $[stit]_i$). Stit grows out from a tradition of modal theories of agency\footnote{Detailed overview of the stit prehistory could be found in \cite{segerberg2009logic}} and originated from a series of seminal articles by Belnap, Perloff and Xu \cite{belnap2001facing}.

In classical stit theory, agent's power to see to it that some state of affairs holds is called a causal ability. The notion ignores any mental attitudes, such as desirability, awareness or purposefulness of the potential action. Casual ability is formalized via a bimodal formula $\Diamond [stit]_i \phi$, i.e. it is possible that agent $i$ sees to it that $\phi$ obtains. Besides the causal abilities, stit theorists study \textit{epistemic} abilities, i.e. agent's power to knowingly see to it that $\phi$. The discussion on  epistemic abilities and their logical form is still open: Horty and Pacuit propose to treat epistemic abilities via action types and to introduce a new modal operator $[kstit]_i \phi$, what reads as \textquotedblleft agent $i$ executes an action type
that she knows to guarantee the truth of $\phi$\textquotedblright:

\begin{quotation}
If the epistemic sense of ability requires that some single action must
be known by $i$ to guarantee the truth of $\phi$, then this must be the
action type, not one of its various tokens  \cite{hort}.
\end{quotation}

While Broersen argue that epistemic ability may be presented simply by adding knowledge operator $K_i \phi$, which satisfies its standard definition from modal epistemic logic. In that case, epistemic ability may be represented as $\Diamond K_i [stit]_i \phi$ -- it is possible that agent $i$ knows that she sees to it that $\phi$ \cite{broersen2011deontic}.

In the paper, we study causal ability as a single monotonic modal operator. In order to present its axiomatisation, we translate standard stit semantics to neighbourhood one and prove correctness and strong completeness w.r.t. a specific class of neighbourhood frames. We show that the mentioned class is modally invariant with BT+AC frames for strandard stit logic. At the end of the paper, some general considerations on getting the same result for epistemic abilities are presented.

\section{Basic theory of \textit{sees to it that}}

The language of "classical" stit logic $\mathbf{L}_{cstit}$ is defined as follows. For a countable set of propositional variables $Var = \{p_1, p_2, \ldots \}$ and a finite set of agents $Ags = \{i_1, i_2, \ldots, i_n \}$:

\[
\phi : = p \: | \: \neg \phi \: | \: \phi \lor \phi \: | \: \Box \phi \: | \: [stit]_i\phi
\]

\noindent where $p \in Var, i \in Ags$. 
Atomic propositions and Boolean connectives have their standard meaning. 

Sees to it that semantics is built upon a branching time structures of the form $\langle Tree, < \rangle$, where $Tree = \{ m_1, m_2, \ldots \}$ is a non-empty set of moments and $<$ is a strict partial order on $Tree$, satisfying \textit{backward linearity} constraint, i.e. for any $m_1, m_2, m_3 \in Tree$, if $m_2 < m_1$ ($m_1$ is preceded by $m_2$) and $m_3 < m_1$, then either $m_3 < m_2$ or $m_2 < m_3$ or $m_3 = m_2$. It is worth noticing that the constraint works only for predecessors: it is totally possible that for some $m_1, m_2, m_3 \in Tree$: $m_2 < m_1$ and $m_2 < m_3$, but $m_3$ and $m_1$ are not $<$-comparable at all.

A maximal set of linearly $<$-ordered moments is a \textit{history}. Each history represents one of the possible  complete courses of events in the world. Formally speaking,  the set of histories is defined as  
\begin{equation*}
\begin{split}
H = \{ h \subseteq Tree \: | \: \forall m_1, m_2 \in h (m_1 < m_2 \lor m_2 < m_1 \lor m_1 = m_2) \land \\  \forall m_3 \not\in h  \exists m_4 \in h (\neg (m_4 < m_3 \lor m_3 < m_4 \lor m_4 = m_3)) \}
\end{split}
\end{equation*}

If some moment $m$ occurs somewhere throughout the history $h$, it can be said that $h$ \textit{passing through} $m$. The set of histories passing through $m$ is denoted as $H_m$. Moment/history pair of the form $m/h$, such that $h \in H_m$, is usually called an \textit{index}. 

\begin{definition}
A branching time (BT) frame is a tuple
\[
\mathcal{F} = \langle Tree, < \rangle 
\]

where $Tree$ is a non-empty set of moments and $<$ is a strict partial ordering on $Tree$, satisfying the backward linearity condition. BT frame is extended to a BT model:

\[
\mathcal{M} = \langle Tree, <,  \nu \rangle
\]

where $\nu: Var \rightarrow 2^{Tree \times H}$ is an evaluation function, mapping each propositional variable to a set of indices, in which the proposition is satisfied.
\end{definition}

In stit theories, agency  is understood as an agents' ability to restrict the set of possible histories to a specific subset. I.e., \textit {agent i sees to it that } $\phi$ at some moment $m$ means that agent $i$ determines that the world will evolve according to one of the histories, satisfying $\phi$ at the moment $m$.

In order to formalize that intuition, BT models should be extended with the finite set of agents $Ags = \{ i_1, i_2, \ldots, i_n \}$ and with a specific function $Choice$, mapping each agent $i$ and moment $m$ into a partition $Choice^m_i$ of $H_m$. Each element of that partition, $X \in Choice^m_i$, represents a set of possible outcomes of $i$'s specific action at $m$. Respectively, $Choice^m_i$ is a collection of such sets. For the sake of simplicity, $X \in Choice^m_i$, such that $h \in X$, is denoted as $Choice^m_i (h)$.

It is important to state that $Choice$ function enjoys the property called \textit{independence of agents}. For all $a, b \in Ags, m \in Tree$: $X \in Choice^m_a \land Y \in Choice^m_b \rightarrow X \cap Y \neq \emptyset$ for any $X, Y \subseteq 2^{H}$. Informally, the property means that every set of agents' choices is consistent: there is no way any agent could execute an action, which would deprive other agents of any choices available for them.

\begin{definition}{BT+AC model}

BT+AC model is defined as a tuple

\[
\mathcal{M} = \langle Tree, <, Ags, Choice, \nu \rangle
\]

\noindent which is BT model extended with a set of agents $Ags$ and $Choice$ function defined as above.

\end{definition}

\begin{definition}{$\mathbf{L}_{cstit}$ semantics}

\noindent $\mathcal{M}, m/h \models p \Leftrightarrow p \in \nu (p)$\\
$\mathcal{M}, m/h \models \neg \phi \Leftrightarrow \mathcal{M}, m/h \not\models \phi$\\
$\mathcal{M}, m/h \models \phi \lor \psi \Leftrightarrow \mathcal{M}, m/h \models \phi \mbox { or } \mathcal{M}, m/h \models \psi$\\
$\mathcal{M}, m/h \models \Box \phi \Leftrightarrow \forall h' \in H_m (\mathcal{M}, m/h' \models \phi)$\\
$\mathcal{M}, m/h \models [stit]_i \phi \Leftrightarrow \forall h' \in Choice^m_i (h) (\mathcal{M}, m/h' \models \phi)$

\end{definition}

\section{Translating stit semantics to neighbourhood semantics: strategic case}

In this section the strategic fragment of $\mathbf{L}_{cstit}$ is surveyed. The fragment is obtained by abandonment of $[stit]_i \phi$ formulas, allowing only statements about agents' abilities. Such statements are treated by a strategic stit modality: $[i]\phi$ stands here for agent $i$ has an ability to achieve $\phi$. Strategic stit modality $[i]$ may be seen as an abbreviation for $\Diamond [stit]_i$. The equivalence of such expressions is shown in the section as well. In order to present strategic stit fragment and its axiomatisation, the classical Krpke semantics for $\mathbf{L}_{cstit}$ is changed to neighbourhood semantics, since $[i]$ operator loses normality\footnote{the seminal paper showing why normal moda logic is inadequate for ability: \cite{kenny1976human}}. It is not hard to notice that $Choice$ function works similarly with neighbourhood functions for classical modal logic: it takes an index and returns a collection of subsets of indices \cite{segerberg1971essay}. 

Another feature worth being observed is redundancy of the main branching time frame component, strict partial order on the set of moments: in  $\mathbf{L}_{cstit}$  there is no formula, whose semantics somehow mentions $<$ relation. To conclude, standard stit model, as it is presented in \cite{hort}, may be redefined as atemporal and neighbourhood one. It  simplifies the task of providing logic, complete w.r.t. corresponding class of frames.

\subsection{Neighbourhood stit: one-shot games and strategic abilities}

The one-shot strategic stit logic's language $\mathbf{L}_{osstit}$ is defined as follows: 
\[
\phi : = p \: | \: \neg \phi \: | \: \phi \lor \phi \: | \: \Box \phi \: | \: [i]\phi \: |\:   [\exists_i] \phi
\]

\noindent where $p \in Var, i \in Ags$. $[i]\phi$ stands for \textquotedblleft agent $i$ is able to see to it that $\phi$\textquotedblright $\:$ (what intuitively corresponds to $\Diamond [stit]_i \phi$  in classical stit), $\langle  \exists_i \rangle \phi$ -- \textquotedblleft agent $i$ is able to execute an action, which does not prevent $\phi$ \textquotedblright $\:$ (or \textquotedblleft agent $i$ could not prevent $\phi$ \textquotedblright $\:$ for $[\exists_i]\phi$), what may be seen as $\Diamond \neg [stit] \neg \phi$ equivalent. Historical necessity modality, as well as Boolean connectives, have their standard meanings. Dual of every modal operator is defined standardly as well. 

It is possible to consider a  one-shot model, i.e. a model for a set of agents simultaneously taking some actions at the unique moment.  
\begin{definition}{One-shot strategic stit (osstit) frames}

\[
\mathcal{F} = \langle W,  \{Choice_i\}_{i \in Ags} \rangle
\]

\begin{itemize}
    \item $W = \{ w_1, w_2, \ldots \}$ is a non-empty set of states. It is suitable to think of $W$ as a set of historically accessible indices, i.e. $\{ m/h \: | \: h \in H_m\}$ for a unique moment $m$.
    \item $Choice_i: W \rightarrow  2^{2^W}$ is a neighbourhood function, defined for every agent. $Choice_i(w)$ is  a set of possible outcomes of $i$'s actions, available for her at $w$.  
    
    \begin{itemize}
        \item Every function $Choice_i$ is monotonic (closed under supersets), does not contain an empty set and contains $W$ itself, i.e. for all $i \in Ags, w \in W$: $\emptyset \not\in Choice_i(w), W \in Choice_i(w)$.
        \item $Choice_i \downharpoonright(w)$ denotes a \textit{non-monotonic core} of neighbourhood $Choice_i(w)$. The non-monotonic core is a set of neighbourhoods, which are not supersets for any other neighbourhoods: $Choice_i \downharpoonright (w) = \{ X \: | \: X \in Choice_i(w) \land \neg \exists Y \in Choice_i(w)(Y \subseteq X)\}$. $Choice_i \downharpoonright(w)$ represents all proper actions, available for $i$ at $w$, without redundant weaker ones. It should be noted that non-monotonic core enjoys the next property (un): $\bigcup Choice_i \downharpoonright (w) = W$. Informally, it says that there is no historically possible state, which could not be an outcome of some proper action of an agent. 
        \item Another crucial property is \textit{independence of agents (ind)}: for all $a,b \in Ags$, $X, Y \subseteq W$, $w \in W$ : $X \in Choice_a(w) \land Y \in Choice_b(w) \rightarrow X \cap Y \neq \emptyset$. Or, interchangeably, $X \in Choice_a(w) \rightarrow W \setminus X \not\in Choice_b(w)$. The property states that every  choice of actions is consistent: there is no way for one agent to take an action, such that another agent would be deprived of some of her choices. 
        
        \item Agents' abilities are historically necessary, i.e. neighbourhoods stay the same over all states \textit{(nec)}: $\forall w, w' \in W \forall X \subseteq W: X \in Choice_i(w) \rightarrow X \in Choice_i(w')$ for any agent $i \in Ags$.
    \end{itemize}
\end{itemize}

\end{definition}

The one-shot strategic stit  model $\mathcal{M} = \langle \mathcal{F}, V \rangle$ extends osstit frame with a standard evaluation function $V: Var \rightarrow 2^W$.

\begin{definition}Neighbourhood strategic stit semantics

\noindent$\mathcal{M}, w \models p \Leftrightarrow w \in V(p)$\\
$\mathcal{M}, w \models \neg \phi \Leftrightarrow \mathcal{M}, w \not\models \phi$\\
$\mathcal{M}, w \models \phi \lor \psi \Leftrightarrow \mathcal{M}, w \models \phi \mbox{ or } \mathcal{M}, w \models \psi$\\
$\mathcal{M}, w \models \Box \phi \Leftrightarrow \forall w' \in W (\mathcal{M}, w' \models \phi)$\\
$\mathcal{M}, w \models [i]\phi \Leftrightarrow \llbracket  \phi \rrbracket \in Choice_i(w)$\\
$\mathcal{M}, w \models [\exists_i] \phi \Leftrightarrow \forall X \in Choice_i \downharpoonright (w) \forall w' \in X (\mathcal{M}, w' \models \phi)$
\end{definition}

As usual, $\llbracket \phi \rrbracket$ abbreviates $\{ w \in W \: | \: \mathcal{M}, w \models \phi \} $. The reader may observe that the following definition for $[i]\phi$ is equivalent to the given above:
\begin{center}
 $\mathcal{M}, w \models [i]\phi \Leftrightarrow \exists X \in Choice_i \downharpoonright(w)(X \subseteq \llbracket \phi \rrbracket)$
 \end{center}

The equivalence of the definitions immediately follows from $Choice_i(w)$ being monotonic. As for the $[\exists_i]\phi$, it can be redefined as a normal modal operator. Consider a binary relation $R_i =  (\bigcup Choice_i \downharpoonright (w))^2$, where $w$ is arbitrary, since $Choice_i \downharpoonright (w)$ be the same for all $w \in W$, what follows from $(nec)$ property.  $R_i$ is an equivalence relation, connecting \textit{all} elements of $i$'s neighbourhoods from non-monotonic core. Then,

\[
\mathcal{M}, w \models [\exists_i] \phi \Leftrightarrow \forall w' \in W (wR_iw' \rightarrow \mathcal{M}, w' \models \phi)
\]

The validity of the given definition can be simply verified.

\begin{table}[]
\begin{center}
    \begin{tabular}{|ll|}
    \hline
    (PL)  & All tautologies of classical propositional logic \\
    (S5$\Box$) & S5 for $\Box$ modality\\
    (S5$[i]$) & MCT4'B logic for $[\exists_i]$\\
    (Incl) & $\Box \phi \rightarrow [i] \phi$\\
    (M) &  $[i](\phi \land \psi) \rightarrow ([i]\phi \land [i] \psi)$\\
    (N) & $ [i]\top$\\
    (D) &  $\neg[i]\bot $\\
    (Pos) & $\Box \phi \equiv [ \exists_i] \phi $\\
    (Nec-A) &  $[i]\phi \rightarrow \Box [i]\phi$\\
    (Ind) & $[1]\phi_1 \land [2]\phi_2 \land \ldots \land [n]\phi_n \rightarrow \Diamond(\phi_1 \land \phi_2 \land \ldots \land \phi_n)$\\
    (RE) & From $\phi \equiv \psi$, infer $[i]\phi \equiv [i]\psi$\\
    (MP) & From $\phi, \phi \rightarrow \psi$, infer $\psi$\\
    \hline
    \end{tabular}
    \end{center} 
    \caption{Axioms for $\mathcal{L}_{osstit}$}
    \label{tab:my_label}
\end{table}

Notice that $\Box$ operator can be viewed as a special case of strategic sees to it that modality, namely, $[\emptyset]$. On semantic level, it could be defined via a special neighbourhood function $Choice_{\emptyset}$, such that $Choice_{\emptyset}: w \mapsto \{ \{W\} \}$ for all $w \in W$. In some occasions it is more suitable to treat $\Box$ as the special case of non-normal modality, while usually it is addressed as an universal S5 modality.

\subsection{Axioms for $\mathcal{L}_{osstit}$}
\begin{theorem}[Soundness and definability]

Recall that the following three properties hold for one-shot strategic stit frames:

\begin{itemize}

  \item  \noindent(ind) $\forall w \in W \forall  X, Y \subseteq W: X \in Choice_a(w) \land Y \in Choice_b(w) \rightarrow X \cap Y \neq \emptyset $\\

   \item  \noindent(nec) $\forall w, w' \in W \forall X \subseteq W: X \in Choice_i(w) \rightarrow X \in Choice_i(w')$\\
   
   \item \noindent (un) $\bigcup Choice_i \downharpoonright (w) = W$
\end{itemize}

\[
\mathcal{F} \models [a] \phi \land [b] \psi \rightarrow \Diamond( \phi \land \psi) \Leftrightarrow \mathcal{F} \models (ind) \footnote{Here and further $\mathcal{F} \models (p)$ means that frame  $\mathcal{F}$ satisfies condition $(p)$}
\]

\begin{proof}
Left to right -- immediately follows from the semantics. Right to left: assume $\mathcal{F} \not\models (ind)$, i.e. $\exists w \in W \exists X, Y \subseteq W: X \in Choice_a(w) \land Y \in Choice_b(w) \land X \cap Y = \emptyset$. Let $V$ be an evaluation such that $\llbracket \phi \rrbracket = X, \llbracket \psi \rrbracket = Y$. Consequently, $\llbracket \phi \rrbracket \in Choice_a(w), \llbracket \psi \rrbracket \in Choice_b(w)$. By semantic definition of  $[a] \mbox{ and } [b]$, $ \langle \mathcal{F}, V \rangle, w \models [a]\phi \land [b] \psi$. Since $X \cap Y = \emptyset, \llbracket \phi \rrbracket \cap \llbracket \psi \rrbracket = \emptyset$, i.e. $ \langle \mathcal{F}, V \rangle \not\models \phi \land \psi$. It follows that $\langle \mathcal{F}, V \rangle, w \models [a]\phi \land [b] \psi \land \neg \Diamond (\phi \land \psi)$ and hence $\mathcal{F} \not\models  [a] \phi \land [b] \psi \rightarrow \Diamond( \phi \land \psi)$.   
\end{proof}

\[
\mathcal{F} \models [i] \phi \rightarrow \Box [i] \phi \Leftrightarrow \mathcal{F} \models (nec)
\]
\begin{proof}
Left to right -- immediately follows from the semantics. Right to left: assume $\mathcal{F} \not\models (nec)$, i.e. $\exists  w, w' \in W \exists X \subseteq W: X \in Choice_i(w) \land X \not\in Choice_i(w')$. Let $V$ be a valuation such that $\llbracket \phi \rrbracket = X$. Then, $\llbracket \phi \rrbracket \in Choice_i(w)$ and $\llbracket \phi \rrbracket \not\in Choice_i(w')$. By semantics of $[i]$, $ \langle \mathcal{F}, V \rangle, w \models [i]\phi$ and $ \langle \mathcal{F}, V \rangle, w' \not\models [i]\phi$. Since $\Box$ is an universal modality and there is $w' \in W$, not satisfying $[i]\phi$, we can conclude that $ \langle \mathcal{F}, V \rangle, w \models \neg \Box [i] \phi$, i.e. $\langle \mathcal{F}, V \rangle, w \models [i] \phi  \land \neg \Box [i] \phi $, hence, $\mathcal{F} \not\models [i] \phi \rightarrow \Box [i] \phi$.   
\end{proof}

\[
 \mathcal{F} \models \Box \phi \equiv [ \exists_i ] \phi \Leftrightarrow  \mathcal{F} \models (un)
\]

\begin{proof}
Left to right -- immediately follows from the semantics. Right to left: Suppose $\mathcal{F} \not \models (un)$, i.e. $ \bigcup Choice_i \downharpoonright (w) \neq W$. The only way that could be true is $\bigcup Choice_i \downharpoonright (w) \subset W$, since every neighbourhood is a collection of subsets of $W$, consequently,  $ \bigcup Choice_i \downharpoonright (w) \setminus W = \emptyset$. Let $ V$ be a valuation function such that $\bigcup Choice_i \downharpoonright (w) = \llbracket  \phi \rrbracket$. Notice that $[ \exists_i ] \phi$ is true at $w$ iff $\bigcup Choice_i \downharpoonright (w) \subseteq \llbracket \phi \rrbracket$ by definition of $ [\exists_i] \phi$. Hence, $ \langle \mathcal{F}, V \rangle, w \models [ \exists_i ] \phi$. By hypothesis, $W \setminus \bigcup Choice_i \downharpoonright (w)$ is not empty, then $\llbracket \neg \phi \rrbracket \neq \emptyset$ as well, hence $ \langle \mathcal{F}, V \rangle \models \neg \Box \phi$. Finally, $\langle \mathcal{F}, V \rangle, w \models [ \exists_i ] \phi \land \neg \Box \phi$, from which it follows that $\mathcal{F} \not\models \Box \phi \equiv [ \exists_i ] \phi$     
\end{proof}
\end{theorem}

Cases for (M), (N), (D), (RE): standard definability results for monotonic neighbourhood frames such that every neighbourhood contains $W$ and does not contain $\emptyset$ \cite{hansen}. (S5) and (Incl) cases: well-established result for modal logic with universal modality \cite{goran}. 

\begin{theorem}[Completeness]

Let $\mathbf{C}$ be a class of all osstit frames, corresponding to Definition 7.Let  $\mathbf{C} \models \phi$ stand for $\mathcal{F} \models \phi$ for every $\mathcal{F} \in \mathbf{C}$. As usual, $\Sigma \models_{\mathbf{C}} \phi$ means that $\phi$ is a semantic consequence of some set of formulas $\Sigma$ in all $\mathbf{C}$-frames. Then,  for arbitrary  $\Gamma \subseteq \mathbf{L}_{osstit}, \phi \in \mathbf{L}_{osstit}$, the following holds: 

\[
\Gamma \models_{\mathbf{C}} \phi \Leftrightarrow \Gamma \vdash_{\mathcal{L}_{osstit}} \phi 
\]
\begin{proof}

\noindent See Appendix.
\end{proof}
\end{theorem}
\subsection{Modal invariance of \textbf{C} and BT+AC frames}
It is important to notice that $Choice_i \downharpoonright$ function lacks one  crucial feature, which BT+AC $Choice$ function enjoys: while $Choice$ returns a \textit{partition} of $H_m$, $Choice_i \downharpoonright$ does not partition $W$. It is easy to find the exact property of partition, which $Choice_i \downharpoonright$ lacks. A collection of subsets $P \subseteq 2^W$ is a partition iff
\begin{enumerate}
    \item $\emptyset \not\in P$
    \item $\bigcup P = W$
    \item $\forall X, Y \in P: X \cap Y = \emptyset$
\end{enumerate}
 While the first two properties are met by definition of osstit frames, the third one is failed. Nevertheless, it is not modally definable: if a class of monotonic neighbourhood frames is modally definable, then it is closed under disjoint unions, generated submodels, bounded morphic images and ultrafilter extensions \cite{hansen}. It is not the case for such class, where  a non-monotonic cores are partitions of the frame's domain. Let a subclass of \textbf{C}-frames, enjoying a property (iii) on its non-monotonic cores, be denoted as $\mathbf{P}$. 
 
 \begin{theorem}$\mathbf{P}$ is not modally definable

 \begin{proof}

 Let $F_1^c = \langle W_1, N_1^c\rangle $ and $F_2^c = \langle W_2, N_2^c \rangle$ be a non-monotonic cores of two monotonic neighbourhood frames $F_1$ and $F_2$.
 \begin{enumerate}
     \item $W_{1} = \{w_1, w_2, w_3, w_4\}, N^{c}_{1}(w_1) = \{\{ w_1, w_2\}, \{ w_3, w_4\} \}$ and for all other worlds, namely, $w_2 ,\ldots, w_4$, $N^{c}_{\emptyset} = N^{c}_{1}(w) = \{ W^{c}_{1} \}$. Obviously, $F_1 \in \mathbf{P}$
     \item $W_2 = \{w_1, w_2, w_3\}, N^c_2(w_1) = \{\{ w_1, w_2\}, \{ w_2, w_3\} \}$ and for all other worlds, namely, $w_2 ,\ldots, w_4$, $N^{c}_{\emptyset} = N^{c}_2(w) = \{W^{c}_{2}\}$. It is not hard to see that $F_2 \not\in \mathbf{P}$, since $\{w_1, w_2\}$, $\{ w_2, w_3\} \in N^c_2(w)$ and $\{w_1, w_2\} \cap \{ w_2, w_3\} \neq \emptyset$
     \item Let $f: W_1^c \rightarrow W^c_2$ be a function, such that
     \begin{align*}
         f: w_1 &\mapsto w_1\\
         f: w_2 &\mapsto w_2\\
         f: w_3 &\mapsto w_3\\
         f: w_4 &\mapsto w_2\\
     \end{align*}
    \item $f$ is a surjective bounded core morphism from $F_1$ to $F_2$. For all $w \in W_1$:
    \begin{enumerate}
        \item If $X \in N^c_1(w)$, then $f[X] \in N^c_2(f(w))$
        \item If $Y \in N^c_2(f(w))$, then there is an $X \subseteq W_1$, such that $f[X] = Y$ and $X \in N^c_1(w)$
    \end{enumerate}
 \end{enumerate}
 
 As it was noted before, $F_1 \in \mathbf{P}, F_2 \not\in \mathbf{P}$ and $F_2$ is a bounded morphic image of $F_1$. Consequently, $ \mathbf{P}$ is not closed under bounded morphic images and hence not modally definable.    
  \end{proof}

 \end{theorem}

\begin{proposition}$\mathcal{L}_{osstit}$ is sound and strongly complete w.r.t. a class of Kripke frames, modally invariant with BT+AC.

 One more important fact is that the non-monotonic core of $\mathbf{P}$-frames may be seen as one-shot \textit{classical} stit frames, i.e. \[\langle \mathbf{m}, Ags, Choice, \nu  \rangle \] \noindent where $\mathbf{m} = \{ m/h \: | \: m \in Tree, h \in H_m \}$ and $Choice$ is a set of $N^c$-functions, indexed by elements of $Ags$. In these models, the semantic definition for ability is
 
 \[
 \mathcal{M}, m/h \models \Diamond [cstit]_i \phi \Leftrightarrow \exists X \in Choice^m_i: X \subseteq \llbracket \phi \rrbracket))
 \]
 
 \noindent which is equivalent to non-monotonic core version of $[i]\phi$ definition in $\mathbf{C}$-frames. As it was showed above, $\mathbf{P}$ differs from $\mathbf{C}$ with the only property, which is modally undefinable. Hence,  $\mathcal{L}_{osstit}$ is sound and strongly complete w.r.t. $\mathbf{P}$-frames as well.
 
 As it was noted in \cite{balb},  BT+AC frames could be replaced by standard Kripke frames without loss of modal invariance. The class of such Kripke frames is nothing but a disjoint union of $\mathbf{P}$ frames, where indices are treated as possible states. From the basic modal logic theory it is known that modally definable class of frames is closed under disjoint unions, consequently, $\mathcal{L}_{osstit}$ is sound and strongly complete w.r.t. Kripke version of BT+AC frames, defined in accordance with \cite{balb}. 
 
\end{proposition}

\begin{definition}Disjoint Union

Let $\mathbf{M} = \{ \langle W_n, \{ Choice^n_i \}_{i \in Ags}, V_n \rangle  \: | \: n \in I  \}$ be a collection of disjoint one-shot strategic stit models. The disjoint union of the models $\biguplus \limits_{n \in I} \mathbf{M}_n = \langle W, \{ Choice_i\}_{i \in Ags}, V \rangle$ is defined as follows:
\begin{enumerate}
    \item $W = \bigcup \limits_{n \in I} W_n$
    \item $V(p) = \bigcup \limits_{n \in I} V_n(p)$
    \item For any $n \in I$, $i \in Ags$, $X \subseteq W$, $w \in W_n$: $X \in Choice_i(w) \Leftrightarrow X \cap W_n \in Choice^n_i(w)$
\end{enumerate}
\end{definition}

\begin{definition}Strategic stit to standard stit translation

\begin{table}[h]
    \centering
    \begin{tabular}{l|l}
    $tr(p)$  & $p$ \\
    $tr(\neg \phi)$ & $\neg tr(\phi)$\\
    $tr(\phi \lor \psi)$ & $tr(\phi) \lor tr(\psi)$\\
    $tr(\Box \phi)$ & $\Box tr(\phi)$\\
    $tr([i]\phi) $ & $\Diamond [stit]_i tr(\phi)$\\
    \end{tabular}
    \caption{Translation from $\mathbf{L}_{osstit}$ to $\mathbf{L}_{cstit}$}
\end{table}
\end{definition}

\begin{theorem}
Let $\mathcal{M} = \langle Tree, <,  Ags, Choice, \nu \rangle $  be a BT+AC a model. Then, for a set of one-shot strategic stit models $ \mathbf{M} = \{ \langle W^m,  \{Choice^m\}_{i \in Ags}, V^m \rangle \}_{n \in Tree}$, such that $W^m = \{m/h \: | \: h \in H_m \}$ for every $m \in Tree$, $Choice^m_i(w) = Choice^m_i \upharpoonright$ \footnote{by $Choice^m_i \upharpoonright$ we mean $Choice^m_i$, closed under supersets}, $V^m(p) = \nu(p)|_{W^m}$, the following holds:

\[
\biguplus \mathbf{M}, m/h \models \phi \Leftrightarrow \mathcal{M}, m/h \models tr(\phi)
\]

\noindent for every  $m/h \in W^m$.
\begin{proof}
Follows directly from Definition 3.7 and Proposition 3.6.  
\end{proof}
\end{theorem}
\section{Conclusion and further research}

We have proposed a fragment of classical stit logic, $\mathcal{L}_{osstit}$,  with a non-normal strategic modality $[i]\phi$. The neighbourhood semantics for the fragment was presented, as well as soundness and strong completeness of $\mathcal{L}_{osstit}$  w.r.t. corresponding class of neighbourhood frames. The latter was obtained by standard method of canonical model construction. Nevertheless, a number of issues are open to further investigation.

It is interesting to consider epistemic extensions of $\mathcal{L}_{osstit}$. It will allow us to reason about epistemic abilities: agent may be able to \textit{knowingly} see to it that $\phi$, i.e. she may be aware of the potential result of an action she is able to execute.   

It may be fruitful to construct such extension by adding a $[K_i]\phi$ modality, what stands for \textquotedblleft agent $i$ has a strategy to \textit{knowingly} see to it that $\phi$\textquotedblright. In comparison with causal strategic stit, the behaviour of the epistemic modality should differ in some aspects.

Consider a case of  epistemically ideal agents, i.e. agents' knowledge is factual, closed under logical consequence and  agents do have positive and negative introspection. Given absolute rationality, it is natural to assume that $[K_i]$ operator  obeys (M), (N) and (D) axioms, just as its \textquotedblleft causal counterpart\textquotedblright  $ \: [i]$. Epistemic abilities are historically necessary, so $(Nec) \: [K_i] \phi \rightarrow \Box [K_i]\phi$ looks intuitive as well.

Obviously, the notion of epistemic ability is stronger than just the causal one. If an agent is able to knowingly see to it that some state of affairs holds, then she is causally able to do it as well: on the level of axioms, it may be represented as  $(Kn-A) \: [K_i] \phi \rightarrow [i]\phi$.

Another way to construct an epistemic version of strategic stit logic is to explicitly introduce knowledge in the language by adding $K_i \phi$ (agent $i$ knows that $\phi$) modal operator for every $i \in Ags$. This step allows to investigate epistemic abilities by treating formulas of the form
\[
[i] K_i \phi
\]

as well as studying knowledge about ability

\[
K_i [i] \phi
\]

and their interplay. Nevertheless, on the level of semantics this extension will require the addition of $\sim_i$ relations to the osstit frames. The desired properties of the indistinguishability relations and their connections with $Choice$ functions are not clear: the most recent discussions on the topic could be found in \cite{hort}, \cite{duijf2021doing}. 

The computational issues left untouched as well. It is known that the general group STIT (i.e. allowing expressions of the form $[stit]_{\Gamma} \phi$, where $\Gamma \subseteq Ags$) without time operators is neither decidable nor finitely axiomatizable in case $|Ags|> 3$ \cite{herzig2008properties}. It is also known that SAT problem for classical atemporal STIT with single-agent modalities is NEXPTIME-complete if $|Ags|> 2$ \cite{balb}. Since $\mathcal{L}_{osstit}$ may be seen as a fragment of the latter, SAT complexity for it is worth investigating.

\newpage
\Appendix
Let $\mathbf{L}$ language be:

\[
\phi : = p \: | \: \neg \phi \: | \: \phi \lor \phi \: | \: [i] \phi \: | \: [\exists_i] \phi \: | \: \Box \phi 
\]
 \noindent for $i \in Ags, p \in Var$. Let $F$ be a frame 

\[
F = \langle W, \{N_i\}_{i \in Ags} \rangle 
\]

\begin{table}[h]
\begin{center}
    \begin{tabular}{|ll|}
    \hline
    (PL)  & All tautologies of classical propositional logic \\
    (S5$\Box$) & S5 for $\Box$ modality\\
    (S5$[\exists_i]$) & MCT4'B logic for $[\exists_i]$\\
    (Incl) & $\Box \phi \rightarrow [i] \phi$\\
    (M) &  $[i](\phi \land \psi) \rightarrow ([i]\phi \land [i] \psi)$\\
    (N) & $ [i] \top$\\
    (D) &  $\neg [i] \bot $\\
    (Pos) & $\Box \phi \equiv [\forall_i] \phi $\\
    (Nec-A) &  $[i]\phi \rightarrow \Box [i] \phi$\\
    (Ind) & $[1] \phi_1 \land [2] \phi_2 \land \ldots \land [n] \phi_n \rightarrow \langle \forall \rangle (\phi_1 \land \phi_2 \land \ldots \land \phi_n)$\\
    (RE) & From $\phi \equiv \psi$, infer $[i] \phi \equiv [i] \psi$\\
    (MP) & From $\phi, \phi \rightarrow \psi$, infer $\psi$\\
    \hline
    \end{tabular}
    \end{center} 
    \caption{Axioms for $\mathcal{L}$}
\end{table}

\begin{table}[h]
    \centering
    \begin{tabular}{l|l}
        (M) & Monotonicity of $N_i$ \\
        (N) & $W \in N_i(w)$\\
        (D) & $\emptyset \not\in N_i(w)$\\
        (T) & $\forall w \in W(w \in N_i(w))$\\
        (4')  & $\forall w \in W \: X \subseteq W (X \in N_i(w) \rightarrow  N^{-1}_i(X) \in N_i(w))$\\
        (B) & $\forall w \in W \: X \subseteq W (w \in X \rightarrow W \setminus( N^{-1}_i(W \setminus X) \in N_i(w) )) $\\
        (Pos)& $\bigcup N_i \downharpoonright (w) = W$\\
        (Nec-A)& $N_i(w) = N_i(w')$\\
        (Ind) & $\forall X \in N_a(w), Y \in N_b(w) (X \cap Y \neq \emptyset)$\\
    \end{tabular}
    \caption{Definability}
\end{table}

Canonicity: a class of frames satisfying (M), (N), (D) is such frames where $N_i$ is monotonic, does not contain $\emptyset$ and   contains $W$. The corresponding logic $MND$ is canonical, hence, strongly complete w.r.t. that class of frames \cite[p. 31, 44]{hansen}. A class of frames satisfying $(S5[\exists_i])$ are such frames where $\bigcup N_i \downharpoonright$ is monotonic, closed under intersections, reflexive, transitive and symmetric. The corresponding logic $(S5[i])$ is canonical hence complete w.r.t. that class of frames \cite[p. 31, 44]{hansen}. $(S5 \Box \bigoplus Incl )$ is strongly complete w.r.t. the class Kripke frames with total accessibility relation \cite[p.100]{pac}.  Therefore, fusion of all three logics satisfying $MND \bigoplus S5([\exists_i]) \bigoplus S5(\Box)$ is canonical hence strongly complete w.r.t. the class of fusions of corresponding frames, i.e. the frames with monotonic $N_i$ functions not containing $\emptyset$ and containing $W$, monotonic reflexive transitive symmetric $\bigcup N_i \downharpoonright$ and universal binary relation $R_i$.

It is left for us to prove that  (Pos), (Nec-A) and (Ind) are  canonical as well.

\begin{definition}Minimal canonical model

For a monotonic modal logic $\mathcal{L}$ and the basic modal language language $\mathbf{L}$, the minimal canonical model is

\[
\mathcal{M}^c = \langle W^c, N^c_i \rangle 
\]

\noindent such that
\begin{enumerate}
    \item $W^c = \{ \Gamma, \Delta, \ldots \}$ is a set of maximal $\mathcal{L}$-consistent sets of $\mathbf{L}$-formulas. By $|\phi| \subseteq W^c$  we denote a proof set of $\phi$, i.e. $|\phi| = \{ \Gamma \: | \: \phi \in \Gamma \}$
    \item $N^c: W^c \rightarrow 2^{2^{W^c}}$ such that $N^c(\Gamma) = \{ |\phi| \: | \: \Box \phi \in \Gamma \}$
    \item $V^c: Var \rightarrow 2^{W^c}$ is an evaluation function such that $\Gamma \in V^c(p)$ iff $p \in \Gamma$
\end{enumerate}
\end{definition}

It is important to notice that monotonic modal logic $M$ is not valid on its \textit{minimal} canonical model, but it is valid on such model, where instead of  $N^c$ there are the same function which is closed under supersets \cite{pac}. We will denote it here and further as $N_c \upharpoonright$ and call it the supplementation of $N^c$.

Now we are ready to construct a canonical model for our logic. 

\begin{definition}Canonical model for $\mathcal{L}$

\[
\mathcal{M}^c = \langle W^c, \{\nu^c_i \upharpoonright\}_{i \in Ags}, V^c \rangle
\]
\begin{enumerate}
    \item $\{ \Gamma, \Delta, \ldots \}$ is a set of maximal $\mathcal{L}$-consistent sets of $\mathbf{L}$-formulas.
    \item $\nu^c_i: W^c \rightarrow 2^{2^{W^c}} $ is a minimal $\mathcal{L}$- canonical neighbourhood function associated with $[i]$ modality. $\nu^c_i \upharpoonright$ is a supplementation for $\nu^c_i$
    \item $V^c: Var \rightarrow 2^{W^c}$ is an evaluation function such that $\Gamma \in V^c(p)$ iff $p \in \Gamma$ 
    \item Notice that if $|\phi| = W^c$, then $|\Box \phi| = W^c$; if $\phi$ is consistent (i.e. $|\phi| \neq \emptyset$), then $|\Diamond \phi| = W^c$. 
    \item As for $[\exists_i]\phi$, it is contained in some $\Gamma \in W^c$ iff $\bigcup \nu^c_i(\Gamma) \subseteq |\phi|$.
\end{enumerate}

\end{definition}

Given the soundness and definability results from Theorem 3.3, the task is to show that  (Pos), (Ind) and (Nec-A) are valid in that canonical model.
\begin{theorem}[(Pos), (Ind) and (Nec-A) are valid on $\mathcal{L}$-canonical model]

\begin{proof}(Pos)

\[\mathcal{M}^c \models \Box\phi \equiv [\exists_i]\phi\]

The left-to-right direction is obvious, so we concentrate on right to left. We prove it by contraposition.

Suppose there is a $\Gamma \in W^c$, such that $\mathcal{M}, \Gamma \models [\exists_i]\phi \land \neg \Box \phi$. Then, $\bigcup \nu^c_i(\Gamma) \subseteq \phi$ (hence, by  definition $[\exists_i]\phi \in \Gamma$ and from the  $\Gamma$ $\mathcal{L}$-consistency of $\Gamma$ it follows that $\Box \phi \in \Gamma$) and $W^c \setminus |\phi| \neq \emptyset$ (by semantics definition of $\neg \Box \phi$). From the latter it is follows that $\neg \phi$ is consistent, i.e. $|\neg \phi| \neq \emptyset$. Then, $|\neg \Box \phi| = W^c$, and if it is so, then $\Gamma \in |\neg \Box \phi|$, i.e. $\neg \Box \phi \in \Gamma$. Hence, $\Gamma$ contains  both $\Box \phi$ and $\neg \Box \phi$, so it is inconsistent, what leads to contradiction.

\end{proof}
%begin{proof}(Alternative: if $MND$ logic contains (Pos), then $\bigcup \nu^c_i(\Gamma) = W^c$ for all $\Gamma$ in $\mathcal{L}$'s canonical model).

%Consider arbitrary $\Gamma \in W^c$. Its minimal neighbourhood $\nu^c_i$ is not empty and do not contain an empty set, since $\Box_i \top, \neg \Box_i \bot \in \Gamma$. So that, $\nu^c_i(\Gamma) = \{|\psi_1|, |\psi_2|, \ldots, |\psi_n| \}$. Consider a formula $\phi:= \bigvee \limits_{1 \leq i \leq n} \psi_n$. $|\phi| = \bigcup \nu^c_i(\Gamma)$ hence $[i]\phi \in \Gamma$. Suppose $W \setminus \bigcup \nu^c_i(\Gamma)  \neq \emptyset$. It is equal to $W \setminus |\phi|  \neq \emptyset$, so that, $\neg \phi$ is consistent and  $\neg [\forall]\phi \in \Gamma$. But since $[i]\phi \in \Gamma$ and $(Pos) \in \Gamma$, $[\forall] \phi \in \Gamma$ too, hence, it is not the case that $W \setminus |\phi| \neq \emptyset$, hence $W \setminus \bigcup \nu^c_i(\Gamma) = \emptyset$ and $W = \bigcup \nu^c_i(\Gamma)$.

%end{proof}

\begin{proof}(Ind)

\[
\mathcal{M}^c \models [a]\phi \land \ldots \land  [b]\psi \rightarrow \Diamond (\phi \land \ldots \land  \psi) 
\]

Suppose that for arbitrary $\Gamma$ it is true that $\mathcal{M}^c, \Gamma \models [a]\phi \land [b]\psi$. By definition, $|\phi| \in \nu^c_a \upharpoonright (\Gamma)$ and $|\psi| \in \nu^c_b \upharpoonright (\Gamma)$. From that, by definition of $\nu^c_i \upharpoonright$, $[a] \phi \land  [b] \psi \in \Gamma$. Then, since $\Gamma$ is $\mathcal{L}$-consistent, $(Ind) \in \Gamma$ and $\Gamma$ is closed under Modus Ponens, so that, $\Diamond (\phi \land \psi) \in \Gamma$. From that, by definition $|\phi \land \psi|\neq \emptyset$, i.e. $\mathcal{M}^c \models \Diamond (\phi \land \psi)$  
\end{proof}
\begin{proof}(Nec-A)

\[
\mathcal{M}^c \models [i] \phi \rightarrow \Box [i] \phi
\]

Suppose that for some $\Gamma$ it is the case that $\mathcal{M}^c, \Gamma \models [i] \phi$, i.e. $|\phi| \in \nu^c_i \upharpoonright (\Gamma)$, then $[i] \phi \in \Gamma$. Since $\Gamma$ is $\mathcal{L}$-consistent, $\Box [i] \phi \in \Gamma$, i.e. $|[i] \phi| = W^c$ hence $\mathcal{M}^c \models \Box \phi$

\end{proof}
\end{theorem}

%% Appendix.
%% Remove the \Appendix command if an
%% appendix is not required.

%Here starts the appendix. If you don't wish an appendix, please remove the \verb|\Appendix| command from the \LaTeX\ file.

%% Bibliography
%% Make sure to use the bibliographystyle aiml22.
%bibliographystyle{aiml22}
%nocite{*}
%bibliography{aiml22.bib}

\end{document}